\documentclass[11pt,a4paper]{article}
\usepackage{jheppub}
\usepackage{hyperref}
\usepackage{mathtext}         
\usepackage{amsmath,amssymb} 
\usepackage{amsfonts}
\usepackage{epsfig}

\newcommand{\be}{\begin{equation}}
\newcommand{\ee}{\end{equation}}
\newcommand{\bea}{\begin{eqnarray}}
\newcommand{\eea}{\end{eqnarray}}
\newcommand{\ba}{\begin{eqnarray*}}
\newcommand{\ea}{\end{eqnarray*}}

\newcommand{\Sp}[1]{{\mbox{Li}}_2\left(#1\right)}
\newcommand{\Fs}[3]{\!\!\left[\begin{array}{c}#1\,;\\#2\,;\end{array}#3\right]}
\newcommand{\Fh}[2]{\,{}_#1F_#2}
\newcommand{\Fx}[2]{\Fs{#1}{#2}{x}}
\newcommand{\Fds}[2]{\Fs{#1}{#2}{\frac{s_{ij}}{4m^2}}}

\renewcommand{\thesection}{\normalsize \arabic {section}}
\begin{document}

\begin{titlepage}


\vspace*{0.2cm}
\begin{center}
{\Large {
Calculation of one-loop integrals for four-photon amplitudes\\
by  functional reduction  method 
\let\thefootnote\relax\footnotetext
{Talk given at the \textit{International Conference on
Quantum Field Theory, High-Energy Physics\\ and Cosmology,}
 Dubna, Russia, July 18-21, 2022.}
}}\\[2 cm]
\end{center}

\begin{center}
{\bf  O.V.~Tarasov}
\vspace{0.5cm}
\\   

\it Joint Institute for Nuclear Research,\\
      141980 Dubna, Russia \\
     E-mail: {\tt otarasov@jinr.ru}
\end{center}

\vspace*{1.0cm}

\noindent
\begin{abstract}

  The method for functional reduction of Feynman
integrals, proposed by the author, is used to calculate
 one-loop integrals corresponding to diagrams with
four external lines. The integrals that emerge from  amplitudes for 
the scattering of light by light, the photon splitting  in an
external field and Delbr\"{u}ck scattering are considered.
For master integrals in  $d$ - dimensions, new analytic results 
 are presented.  For $d=4$, these integrals are given by compact
expressions in terms of logarithms and dilogarithms.
\end{abstract}

\end{titlepage}

\section{Introduction}
In ref.~\cite{Tarasov:2022clb} a method for functional
reduction of one-loop integrals with arbitrary masses and external
momenta was proposed.
In the present article I will describe the application of this method 
for  calculating one-loop integrals that arise  when  computing radiative 
corrections to important physical processes.
As an example, I chose to calculate integrals required for 
radiative corrections to the amplitudes of scattering of light by
light,  photon  splitting in an external field and Delbr\"{u}ck
scattering.
First calculations of radiative corrections to these
processes were presented  in refs.
\cite{Karplus:1950zz}--\cite{Constantini:1971}. 
Radiative corrections to the amplitude of the photon splitting in an external
field were considered in refs. \cite{Shima:1966cmi}, \cite{Baier:1974ga}.
Electroweak radiative corrections to the process of 
photon-photon scattering were investigated in refs.
\cite{Jikia:1993tc}, \cite{Gounaris:1999gh}.

The study of the scattering of light by
light is an important part of the experimental  program at the LHC.
The main experiment in this study is collisions of lead ions.
The first results of these experiments were reported in refs.
\cite{ATLAS:2017fur}, \cite{CMS:2018erd}.

The aim of this paper is to calculate  integrals that arise when computing
 amplitudes for the processes with four external photons. 
Note that integrals of this type can be used to calculate
radiative corrections to other processes, as well as to calculate
diagrams with five and more external lines that can be reduced
to the considered integrals.

\section{Integrals and the functional reduction formula}

We will consider the calculation of integrals of the following type
\begin{equation}
I_4^{(d)}(m_1^2,m_2^2,m_3^2,m_4^2; 
s_{12},s_{23},s_{34},s_{14},s_{24},s_{13})
=\frac{1}{i \pi^{d/2}} \int \frac{d^dk_1}{D_1D_2D_3D_4},
\end{equation}
where
\begin{equation}
D_j=(k_1-p_j)^2-m_j^2+i\epsilon,~~~~~~~~s_{ij}=(p_i-p_j)^2.
\label{product4P}
\end{equation}
In what follows, we will omit the small imaginary term $i\epsilon$,
implying that all the masses contain it. 
Figure 1 shows diagrams corresponding to the integrals
that we  consider in this paper.
\vspace{0.5cm}
\begin{figure}[h]
\begin{center}
\includegraphics[scale=0.9]{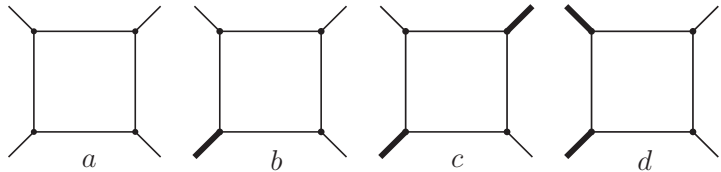}
\end{center}
\caption{
Diagrams corresponding to scattering of light by  light,
photon splitting and Delbr\"{u}ck scattering.
The thick lines correspond to the off-shell photon. }
\end{figure}

To calculate the integral $I_4^{(d)}$, we will need
integrals with fewer propagators:
\begin{eqnarray}
&&
I_3^{(d)}(m_1^2,m_2^2,m_3^2; s_{23},s_{13},s_{12})
=\frac{1}{i \pi^{d/2}} \int \frac{d^dk_1}{D_1D_2D_3},
\\
&&
I_2^{(d)}(m_1^2,m_2^2; s_{12})
=\frac{1}{i \pi^{d/2}} \int \frac{d^dk_1}{D_1D_2},
\\
&&I_1^{(d)}(m_1^2)
=\frac{1}{i \pi^{d/2}} \int \frac{d^dk_1}{D_1}.
\end{eqnarray}

The formula for  functional reduction of the integral $I_4^{(d)}$ with
arbitrary masses and kinematic variables
\cite{Tarasov:2022clb} can be represented as, 
\begin{eqnarray}
&&I_4^{(d)}(m_1^2,m_2^2,m_3^2,m_4^2; s_{12},s_{23},s_{34},s_{14},s_{24},s_{13})
\nonumber \\
&&~~~~~~~~~~~~~~~~~~~~~~~~~~~~
=
\sum_{k=1}^{24} Q_k 
{ B_4^{(d)}(}R^{(k)}_{1},R^{(k)}_{2},R^{(k)}_{3},R^{(k)}_{4}),
\label{I4viaB4}
\end{eqnarray}
where $R^{(k)}_{j}$ are the ratios of the modified Cayley determinant
 to the Gram determinant, $Q_k$ are products of the ratios
of polynomials in the squared masses and kinematic variables $s_{ij}$.
The function $B_4^{(d)}$ is defined as follows
\begin{eqnarray}
&&{ B_4^{(d)}(R_1,R_2,R_3,R_4)}  \nonumber \\
&&
~~~=I_4^{(d)}(R_1,R_2,R_3,R_4; R_2\!-\!R_1,R_3\!-\!R_2,R_4\!-\!R_3,
R_4\!-\!R_1, R_4\!-\!R_2,R_3\!-\!R_1)
\nonumber \\
&&~~~=\Gamma\left(4-\frac{d}{2}\right)
R_1^{\frac{d}{2}-4}
\int_0^1\int_0^1\int_0^1x_1^2x_2~h_4^{\frac{d}{2}-4} dx_1dx_2dx_3,
\label{int_rep_B4}
\end{eqnarray}
\begin{equation*}
h_4=1-z_1 x_1^2- z_2 x_1^2x_2^2
-z_3 x_1^2x_2^2x_3^2.
\end{equation*}
The variables $z_j$ are simple combinations of $R_i$
\begin{equation}
z_1=1-\frac{R_{2}}{R_{1}},~~~~z_2=\frac{R_{2}-R_{3}}{R_{1}},
~~~~z_3=\frac{R_{3}-R_4}{R_{1}}.
\end{equation}
An explicit formula for the functional reduction of the integral $I_4^{(d)}$
is presented in \cite{Tarasov:2022clb}.

\section{Integral $I_4^{(d)}$ for the  light by light  scattering  amplitude }
The integral required for computing the 
light by light scattering  amplitude corresponds to the following 
values of kinematic variables and masses
\begin{equation}
s_{12}=s_{23}=s_{34}=s_{14}=0,~~~~~~m_j^2=m^2, ~~~~j=1,...,4.
\label{LbyL}
\end{equation}
Inserting these values  in the final formula for functional
reduction of the integral $I_4^{(d)}$, given in
\cite{Tarasov:2022clb}, we get
\begin{eqnarray}
&&I_4^{(d)}(m^2,m^2,m^2,m^2; 0,0,0,0,s_{24},s_{13})
=
\nonumber \\
&&~~~~~~~~~~~~~~~
\frac{\tilde{s}_{13}}{(\tilde{s}_{24}+\tilde{s}_{13})} L^{(d)}(\tilde{s}_{13},M_2^2,m^2)
+\frac{\tilde{s}_{24}}{(\tilde{s}_{24}+\tilde{s}_{13})} L^{(d)}(\tilde{s}_{24},M_2^2,m^2),
\label{boxLbyL}
\end{eqnarray}
where
\begin{eqnarray}
&&L^{(d)}(\tilde{s}_{ij},M_2^2,m^2)=
\nonumber \\
&&~~~~~~
I_4^{(d)}\left( m^2-M_2^2, m^2,m^2- \tilde{s}_{ij},m^2;M^2_2,-\tilde{s}_{ij},\tilde{s}_{ij},
M^2_2 ,0,M^2_2-\tilde{s}_{ij} \right),~~~~~
\end{eqnarray}
\begin{equation}
M^2_2=\frac{\tilde{s}_{24}\tilde{s}_{13}}{(\tilde{s}_{24}+\tilde{s}_{13})},~~~~~~~\tilde{s}_{ij}=\frac{s_{ij}}{4}.
\end{equation}
The resulting formula (\ref{boxLbyL}) represents the integral $I_4^{(d)}$ depending on three variables
in terms of integrals also depending on three variables.
However, calculating the integral $L^{(d)}$ is simpler than calculating
the original integral.
We will consider different methods of calculating $L^{(d)}$.

 It turns out that the recurrence equation with respect to
 $d$ provides an easy way to derive a compact expression
 for the function $L^{(d)}$.
  Such an equation
for the integral $L^{(d)}(\tilde{s}_{ij},M_2^2,m^2)$ has the form
\cite{Tarasov:2022clb}:
\begin{equation}
(d-3)L^{(d+2)}(\tilde{s}_{ij},M_2^2,m^2)
=-2 m^2_0 L^{(d)}(\tilde{s}_{ij},M_2^2,m^2)-I_3^{(d)}(m^2;\tilde{s}_{ij}),
\label{dim_rec_os}
\end{equation}
where
\begin{equation}
I_3^{(d)}(m^2;\tilde{s}_{ij})=
I_3^{(d)}(m^2,m^2-\tilde{s}_{ij},m^2; \tilde{s}_{ij},0,-\tilde{s}_{ij}),
\end{equation}
\begin{equation}
m_0^2= m^2-M_2^2. 
\end{equation}
The solution of the equation (\ref{dim_rec_os}) can be represented
as
\begin{eqnarray}
&&
L^{(d)}(\tilde{s}_{ij},M^2_2,m^2)
\nonumber \\
&&~~~~~~~
=
\frac{(m_0^2)^{\frac{d}{2}} ~c_4^{(d)}(\tilde{s}_{ij},M^2_2,m^2)}
{\Gamma\left(\frac{d-3}{2}\right) \sin \frac{\pi d}{2}}
-\frac{1}{2m_0^2}\sum_{r=0}^{\infty}
\frac{\left(\frac{d-3}{2}\right)_r}{(-m_0^2)^r}
I_3^{(d+2r)}(m^2;\tilde{s}_{ij}),
\label{solu_dim_rec}
\end{eqnarray}
where
 $c_4^{(d+2)}(\tilde{s}_{ij},M^2_2,m^2)=c_4^{(d)}(\tilde{s}_{ij},M^2_2,m^2)$
is an arbitrary periodic function of the parameter $d$.
Using the method of ref. \cite{Tarasov:2015wcd}, we obtained for the integral
$I_3^{(d)}(m^2;\tilde{s}_{ij})$ a simple functional relation
\begin{equation}
I_3^{(d)}(m^2;\tilde{s}_{ij})=
I_3^{(d)}(m^2,m^2,m^2; 4\tilde{s}_{ij},0,0),
\end{equation}
which reduces this integral to the  well-known result \cite{Boos:1990rg}
\begin{eqnarray}
I_3^{(d)}(m^2;\tilde{s}_{ij})
= 
- \frac{(m^2)^{\frac{d}{2}-3}}{2} \Gamma\left(3-\frac{d}{2}\right)
\Fh32\Fds{1,1,3-\frac{d}{2}}{2,\frac{3}{2}}.
\label{I3small_Mom}
\end{eqnarray}
We find the function $c_4^{(d)}(\tilde{s}_{ij},M^2_2,m^2)$ as a solution 
of a differential equation that can be obtained, for example,
from a differential equation 
for the integral $L^{(d)}(\tilde{s}_{ij},M_2^2,m^2)$
\begin{eqnarray}
&&\tilde{s}_{ij}\frac{\partial}{\partial \tilde{s}_{ij}}
L^{(d)}(\tilde{s}_{ij},M_2^2,m^2) = -L^{(d)}(\tilde{s}_{ij},M_2^2,m^2)
\nonumber \\
&&~~~~~~~~~~~~~~~
+\frac{(d-3)}{8(M_2^2-\tilde{s}_{ij})(M_2^2-m^2)}I^{(d)}_2(m^2-M_2^2,m^2;M_2^2)
\nonumber \\
&&~~~~~~~~~~~~~~~
 -\frac{(d-3)}{8(M_2^2-\tilde{s}_{ij})(\tilde{s}_{ij}-m^2)}
 I_2^{(d)}(m^2-\tilde{s}_{ij},m^2;\tilde{s}_{ij})
\nonumber \\
&&~~~~~~~~~~~~~~~
 +\frac{(d-2)}{16m^2(\tilde{s}_{ij}-m^2)(M_2^2-m^2)}I_1^{(d)}(m^2).
\label{proI4_S13}
\end{eqnarray}
This equation was obtained under the assumption that $\tilde{s}_{ij}$ and $M_2^2$
are independent variables.
Substituting the solution (\ref{solu_dim_rec}) into the equation (\ref{proI4_S13}),
we get a simple differential equation
\begin{equation}
\tilde{s}_{ij}\frac{\partial c_4^{(d)}(\tilde{s}_{ij},M_2^2,m^2)}
{\partial \tilde{s}_{ij}}+c_4^{(d)}(\tilde{s}_{ij},M_2^2,m^2)=0.
\label{dif_equ4c4}
\end{equation}
The solution of this equation is
\begin{equation}
c_4^{(d)}(\tilde{s}_{ij},M_2^2,m^2)
= \frac{\tilde{c}_4^{(d)}(M_2^2,m^2)}{\tilde{s}_{ij}},
\end{equation}
where $\tilde{c}_4^{(d)}(M_2^2,m^2)$ is the integration constant
of the differential equation (\ref{dif_equ4c4}).
Using the boundary value of the integral $L^{(d)}$ at $\tilde{s}_{ij}=0$, we get
\begin{equation}
c_4^{(d)}(\tilde{s}_{ij},M_2^2,m^2) = 0.
\label{c4zero}
\end{equation}

For the hypergeometric function from the equation (\ref{I3small_Mom}),
we used the following representation \cite{rainville1960special}
\begin{eqnarray}
\Fh32\Fx{1,1,3-\frac{d}{2}}{2,\frac{3}{2}}=\frac{-\Gamma\left(\frac32\right)}
{x  \Gamma\left(3-\frac{d}{2}\right) \Gamma\left(\frac{d-3}{2}\right)}
\int_0^1 dz z^{\frac{2-d}{2}}(1-z)^{\frac{d-5}{2}}
\ln(1-xz).~~~~
\label{I3intrepr}
\end{eqnarray}
Employing this representation and taking into account equation (\ref{c4zero}), 
the solution (\ref{solu_dim_rec}) can be written as
\begin{eqnarray}
L^{(d)}(\tilde{s}_{ij},M_2^2,m^2) = -
\frac{\pi^{\frac12}(m^2)^{\frac{d}{2}-3}}
{8\tilde{s}_{ij}\Gamma\left(\frac{d-3}{2}\right)}
\int_0^1 \frac{dz ~z^{2-\frac{d}{2}}(1-z)^{\frac{d-5}{2}}}
{1-z\frac{M_2^2}{m^2}}\ln\left(1-z\frac{\tilde{s}_{ij}}{m^2}\right).
\label{I4_LbyL}
\end{eqnarray}
Substituting this expression into the equation (\ref{boxLbyL}), we get
\begin{eqnarray}
&&I_4^{(d)}(m^2,m^2,m^2,m^2;  0,0,0,0,s_{24},s_{13})
=
-\frac{\pi^{\frac12}(m^2)^{\frac{d}{2}-3}}
{2(s_{24}+s_{13})\Gamma\left(\frac{d-3}{2}\right)}
~~~~~~~~~~~
\nonumber \\
&& ~~~~~~~~~~\times
\int_0^1 \frac{dz ~z^{2-\frac{d}{2}}(1-z)^{\frac{d-5}{2}}}
{1-z\frac{M_2^2}{m^2}}
\left[ \ln\left(1-z\frac{s_{13}}{4m^2}\right)+
\ln\left(1-z\frac{s_{24}}{4m^2}\right) \right].~~~
\label{SumI_LbL}
\end{eqnarray}
This is the simplest representation known so far
for this integral. Note that for $d=4$ in ref.
\cite{Davydychev:1993ut}, this integral was obtained
in terms of the hypergeometric Appell function $F_3$,
which was expressed as a two-fold integral.

The representation (\ref{I4_LbyL}) is convenient for 
a series expansion in $\varepsilon =(4-d)/2$. For $d=4$, we have
\begin{eqnarray}
L^{(4)}(\tilde{s}_{ij},M_2^2,m^2)=
\frac{-1}{8M_2^2{s}_{ij}\beta_{2}}
\left[
\ln^2\frac{\beta_{ij}+1}{\beta_{2}-\beta_{ij}}
-\frac12 \ln^2\frac{\beta_{2}-\beta_{ij}}{\beta_{2}+\beta_{ij}}
-\ln \frac{\beta_{ij}-1}{\beta_{ij}+1}
\ln\frac{\beta_{2}-\beta_{ij}}{\beta_{2}+\beta_{ij}}
\right.
\nonumber \\
\nonumber \\
\left.
+\frac{\pi^2}{3}
+2\Sp{\frac{\beta_{ij}-\beta_{2}}{\beta_{ij}+1}}
-2\Sp{\frac{\beta_{ij}-1}{\beta_{2}+\beta_{ij}}}
+\Sp{\frac{\beta_{ij}^2-1}{\beta_{ij}^2-\beta_{2}^2}}
\right].~~~~~~~~
\label{I_LbL}
\end{eqnarray} 
where
\begin{equation}
\beta_{ij} \equiv \sqrt{1- \frac{m^2}{\tilde{s}_{ij}}},~~~~(ij=13,24) \; \;~~~~~~~~~~~
\beta_{2} \equiv \sqrt{1-\frac{m^2}{M_2^2}}. \;
\label{Betass}
\end{equation}
Using (\ref{I_LbL}), from the equation (\ref{boxLbyL}),
we get
\begin{eqnarray}
&&I_4^{(4)}(m^2,m^2,m^2,m^2;  0,0,0,0,s_{24},s_{13})
\nonumber \\ [0.3 cm]
&& = \frac{1}{8\tilde{s}_{13}\tilde{s}_{24} \beta_{2}}
 \left\{ 2 \ln^2
    \left( \frac{\beta_{2}+\beta_{13}}{\beta_{2}+\beta_{24}} \right)
+ \ln \left( \frac{\beta_{2}-\beta_{13}}{\beta_{2}+\beta_{13}} \right) \;
     \ln \left( \frac{\beta_{2}-\beta_{24}}{\beta_{2}+\beta_{24}} \right)
   - \frac{\pi^2}{2} \right. \hspace{1cm}
\nonumber \\ 
&& \left.
  +\sum_{i=13,24}^{} \left[
    2 \Sp{\frac{\beta_{i} -1}{\beta_{2}+\beta_i}}
   -2 \Sp{-\frac{\beta_{2}-\beta_i}{\beta_{i} +1}}
  - \ln^2
    \left( \frac{\beta_{i} +1}{\beta_{2}+\beta_i} \right)
  \right] \right\}.~~~~~~~~~
\label{I4eps0} 
\end{eqnarray}
In the sum of two $L^{(d)}$ the third terms with $\rm Li_2$ from (\ref{I_LbL}) 
cancel.
The expression (\ref{I4eps0}) agrees with the result obtained in
ref. \cite{Davydychev:1993ut}.

The function $L^{(d)}$ can be obtained as a solution to the equation (\ref{proI4_S13}).
For $d=4$, the solution of this equation has the form
\begin{eqnarray}
&&L^{(4)}(\tilde{s}_{ij},M_2^2,m^2) =
\frac{1}{8s_{ij}M_2^2 \beta_2}
\Biggl[F(y_{ij},y_2)
-2{\rm Li_2}\left(1- y_2\right)
\Biggr],~~~~~~
\label{best_I4LbL}
\end{eqnarray}
where
\begin{equation}
F(y_{ij},y_k)={\rm Li}_2\left(1-y_{ij}y_k\right)+
{\rm Li}_2\left(1-\frac{y_k}{y_{ij}}\right)+\frac12 \ln^2y_{ij},
\end{equation}
\begin{equation}
y_{ij}=\frac{\beta_{ij}-1}{\beta_{ij}+1},~~~~~~y_{2}=\frac{\beta_{2}-1}{\beta_{2}+1}.
\end{equation}  
Note that the expression (\ref{best_I4LbL}) is invariant as with respect 
to the replacement of $y_{ij}\rightarrow 1/y_{ij}$ and 
with respect to the replacement 
$y_{2}\rightarrow 1/y_{2}$.

Substituting (\ref{best_I4LbL}) into the formula (\ref{boxLbyL}), we get
\begin{eqnarray}
&&I_4^{(d)}(m^2,m^2,m^2,m^2; 0,0,0,0,s_{24},s_{13})
\nonumber \\
&&~~~~~
=\frac{1}{2 s_{13}s_{24}\beta_2}
\Biggl[
F(y_{13},y_2)+F(y_{24},y_2)-4{\rm Li_2}\left(1- y_2\right)
\Biggr].~~~
\label{five_li}
\end{eqnarray}

Series representations of $L^{(d)}(\tilde{s}_{ij},M_2^2,m^2)$
can be obtained from the integral representations (\ref{int_rep_B4}), (\ref{I4_LbyL}).

\section{
Integral $I_4^{(d)}$ for the photon splitting amplitude}

In this section, we turn to the
 integral with one external line off-shell, which is 
associated with the diagram $b)$ in Figure 1.
This integral corresponds to the kinematics
\begin{equation}
s_{23}=s_{34}=s_{14}=0,~~~~~~m_i^2=m^2.
\end{equation}
Using the final formula for functional reduction of
$I_4^{(d)}$ from ref. \cite{Tarasov:2022clb}, we get
\begin{eqnarray}
&&(s_{12}-s_{24}-s_{13})I_4^{(d)}(m^2,m^2,m^2,m^2,s_{12},0,0,0,s_{24},s_{13})
\nonumber \\
&& \nonumber \\
&&~~~~= s_{12}
L^{(d)}(\tilde{s}_{12},M_3^2,m^2)-s_{13}L^{(d)}(\tilde{s}_{13},M_3^2,m^2)
-s_{24}
L^{(d)}(\tilde{s}_{24},M_3^2,m^2).~~~~
\end{eqnarray}
where 
\begin{equation}
M_3^2=\frac{s_{24}s_{13}}{4(s_{13}+s_{24}-s_{12})}.
\end{equation}
For $d=4$, we have
\begin{eqnarray}
&&I_4^{(4)}(m^2,m^2,m^2,m^2,s_{12},0,0,0,s_{24},s_{13})
\nonumber \\
&& \nonumber \\
&&~~~~~~~~ =\frac{
F(y_{13},y_3)+F(y_{24},y_3)-F(y_{12},y_3)
-2 {\rm Li}_2(1-y_3)}{2{s}_{24}{s}_{13}\beta_3},
\end{eqnarray}
where
\begin{equation}
y_3=\frac{\beta_3-1}{\beta_3+1},~~~~~\beta_3=\sqrt{1-\frac{m^2}{M_3^2}}.
\end{equation}

\section{
Integrals $I_4^{(d)}$ for the  Delbr\"{u}ck scattering amplitude
}
We now turn to the more complicated integrals corresponding to the
diagrams $c)$ and $d)$ in Figure 1.

Inserting the values
of masses and kinematic variables for the integral represented by
the diagram $c)$,
\begin{equation}
s_{23}=s_{14}=0,~~~~~m_k^2=m^2, ~~~k=1,...,4,
\end{equation} 
in the final formula for the functional reduction of $I_4^{(d)}$ from
ref. \cite{Tarasov:2022clb}, we get
\begin{eqnarray}
&&(s_{12}-s_{13}-s_{24}+s_{34})
I_4^{(d)}(m^2,m^2,m^2,m^2;s_{12},0,s_{34},0,s_{24},s_{13})
\nonumber \\
&&~~~~~~~~~~~~~~~~=
 s_{12}L^{(d)}(\tilde{s}_{12},M_4^2,m^2)
-s_{13}L^{(d)}(\tilde{s}_{13},M_4^2,m^2)
\nonumber \\
&&~~~~~~~~~~~~~~~~
-s_{24}L^{(d)}(\tilde{s}_{24},M_4^2,m^2)
+s_{34}L^{(d)}(\tilde{s}_{34},M_4^2,m^2),~~~~~
\end{eqnarray}
where
\begin{eqnarray}
&& M^2_4=\frac{s_{12}s_{34}-s_{13}s_{24}}{4(s_{12}-s_{13}-s_{24}+s_{34})}.
\end{eqnarray}

At $d=4$, we obtain
\begin{eqnarray}
&&
I_4^{(4)}(m^2,m^2,m^2,m^2;s_{12},0,s_{34},0,s_{24},s_{13})
\nonumber \\
&& \nonumber \\
&&~~~~~~~~~~~
=\frac{F(y_{13},y_4)+F(y_{24},y_4)-F(y_{34},y_4)-F(y_{12},y_4)
}
{2({s}_{13}{s}_{24}-{s}_{12}{s}_{34})\beta_4},~~~
\end{eqnarray}
where
\begin{equation}
y_4=\frac{\beta_4-1}{\beta_4+1},~~~~~\beta_4=\sqrt{1-\frac{m^2}{M_4^2}}.
\end{equation}

Another integral contributing to Delbr\"{u}ck
scattering cross section is represented by the diagram $d)$ in Figure 1.
The kinematics corresponding to this diagram reads
\begin{equation}
s_{34}=s_{14}=0,~~~~m_k^2=m^2,~~~~k=1,...4.
\end{equation}
Substituting these values into the final formula for functional
reduction \cite{Tarasov:2022clb}, we obtain
\begin{eqnarray}
&&d_1d_2
I_4^{(d)}(m^2,m^2,m^2,m^2;s_{12},s_{23},0,0,s_{24},s_{13})
\nonumber \\
&&=
\frac12 d_1 s_{13} s_{24}I_4^{(d)}(m_0^2,m^2,m^2-\tilde{s}_{13},m^2)
-\frac12 d_1 s_{23} s_{24}I_4^{(d)}(m_0^2,m^2,m^2-\tilde{s}_{23},m^2)
\nonumber \\
&&
-~\frac12 d_1 s_{12}s_{24} I_4^{(d)}(m_0^2,m^2,m^2-\tilde{s}_{12},m^2)
+d_1 s_{24}^2I_4^{(d)}(m_0^2,m^2,m^2-\tilde{s}_{24},m^2)
\nonumber \\
&&
+~\frac12 n_1 s_{12}(s_{12}-s_{13}-s_{23})
I_4^{(d)}(m_0^2,\tilde{m}^2_0,m^2-\tilde{s}_{12},m^2)
\nonumber \\
&&-~ 
\frac12 n_1 s_{13}(s_{12}-s_{13}+s_{23})
I_4^{(d)}(m_0^2,\tilde{m}^2_0,m^2-\tilde{s}_{13},m^2)
\nonumber \\
&&-~
\frac12 n_1 s_{23}(s_{12}+s_{13}-s_{23})
I_4^{(d)}(m_0^2,\tilde{m}^2_0,m^2-\tilde{s}_{23},m^2),
\label{I4fig_d}
\end{eqnarray}
where
\begin{equation}
m_0^2 = m^2-\tilde{s}_1,
~~~~~~\tilde{m}^2_0=m^2-\tilde{s}_2,
~~~~~~\tilde{s}_1 = \frac{s_{13}s_{24}^2}{4 d_2},
~~~~~~\tilde{s}_2=-\frac{s_{12}s_{13}s_{23}}{d_1},
\end{equation}
\begin{eqnarray}
&&n_1=2s_{12}s_{23}-s_{12}s_{24}+s_{13}s_{24}-s_{23}s_{24}, \nonumber \\
&&d_1=s_{12}^2+s_{13}^2+s_{23}^2-2 s_{12}s_{13}-2 s_{12}s_{23}-2s_{13}s_{23},
\nonumber \\
&&
d_2=s_{12}s_{23}-s_{12}s_{24}+s_{13}s_{24}-s_{23}s_{24}+s_{24}^2.
\end{eqnarray}
The first four integrals in the equation (\ref{I4fig_d}) are 
expressed in terms of the   function $L^{(d)}$. The three remaining integrals 
$I_4^{(d)}(m_0^2,\tilde{m}^2_0,m^2-\tilde{s}_{ij},m^2)$, 
can be calculated using the formula from ref. \cite{Tarasov:2022clb}
\begin{eqnarray}
&&I_4^{(d)}(r_{1234},r_{234},r_{34},r_4)
 =  
\int_0^1\int_0^1\!\int_0^1\frac{\Gamma\left(4-\frac{d}{2}\right)
x_1^2x_2~ dx_1dx_2dx_3 }
 {\left[
 a-b x_1^2-c x_1^2x_2^2 -e x_1^2x_2^2x_3^2
\right]^{4-\frac{d}{2}}},~~~~~~~
\label{I4param}
\end{eqnarray}
where
\begin{equation}
a=r_{1234},~~~~~b=r_{1234}-r_{234},~~~~~c=r_{234}-r_{34},
~~~~~e=r_{34}-r_4.
\end{equation}
For $d=4$, the integral (\ref{I4param}) can be calculated as a
combination
of functions $L^{(4)}$ with various arguments. The detailed  derivation
of the result will be described in an expanded version of this article.

\section{
Conclusions}
Our results clearly indicate that  the application
of the functional reduction method makes it possible to reduce
 complicated integrals to simpler integrals.
Even if the number of variables in the integrals resulting from 
applying 
the functional reduction is the same as in the original integral,
these integrals are simpler than the original integral.
In general, integrals depending on more than four variables
 are reduced to a combination of integrals with four or
fewer variables, which are also simpler than the original integral.

The fact that the results for the  integrals discussed in this paper are expressed in
terms of the same function $L^{(d)}$ may be useful
for  improving the accuracy and   efficiency 
of  calculating radiative corrections.
The integral representation (\ref{I4_LbyL}) can significantly simplify 
the calculation of  higher order terms in the 
$\varepsilon$ expansion of integrals considered in the article.

\end{document}